\documentclass[aps,prd,onecolumn,groupedaddress,showpacs,nofootinbib,amssymb]{revtex4}
\usepackage[dvips]{graphicx}
\usepackage{amssymb}
\usepackage{amsmath}
\usepackage{graphicx,,color}
\usepackage{amsfonts}
\usepackage{bm}
\usepackage{cancel}
\usepackage{comment}

\newcommand\be{\begin{equation}}
\newcommand\ee{\end{equation}}

\allowdisplaybreaks[4]

\begin{document}

\tolerance=5000

\title{Non-minimally Coupled Einstein-Gauss-Bonnet Gravity with Massless Gravitons: The Constant-roll Case}
\author{V.K.~Oikonomou,$^{1,2,3}$\,\thanks{v.k.oikonomou1979@gmail.com}F.P.
Fronimos,$^{1}$\,\thanks{fotisfronimos@gmail.com}} \affiliation{
$^{1)}$ Department of Physics, Aristotle University of
Thessaloniki, Thessaloniki 54124,
Greece\\
$^{2)}$ Laboratory for Theoretical Cosmology, Tomsk State
University of Control Systems and Radioelectronics, 634050 Tomsk,
Russia (TUSUR)\\
$^{3)}$ Tomsk State Pedagogical University, 634061 Tomsk,
Russia\\}

\tolerance=5000

\begin{abstract}
In this letter we study the behavior of non-minimally coupled
Einstein-Gauss-Bonnet gravity theories with the constant-roll
condition. Recalling the results of the striking GW170817 event,
we demand that the velocity of the gravitational waves is equated
to unity in natural units, meaning that $c_T^2=1$. This is a
powerful restriction since it leads to a decrease in the degrees
of freedom and subsequently reveals a connection between the
scalar functions of the theory which presumably have different
origins. In this framework, we shall assume that a scalar
potential is present and can be extracted easily from the
equations of motion by simply designating the scalar coupling
functions. Obviously, a different approach is feasible but such
choice will prove to be extremely convenient. Afterwards, we
impose certain approximations in order to facilitate our study.
Each assumption is capable of producing different phenomenology so
a summary of all the possible configurations for Hubble's
parameter, its derivative and the scalar potential along with the
corresponding assumptions are present at the end of the paper. We
show that compatibility under the constant-roll assumption can be
achieved for a variety of model functions and different
approaches, although one must always be aware of the imposed
approximations since the possibility of a model producing viable
results while simultaneously violating even a single approximation
exists. Finally, in the end, a new formalism which leads to a
plethora of convenient coupling functions according to the readers
choice is presented. Utilizing such formalism may lead to new
cases for coupling functions which have not been used yet, but are
in fact able of producing viable phenomenology for the
inflationary era.
\end{abstract}

\pacs{04.50.Kd, 95.36.+x, 98.80.-k, 98.80.Cq,11.25.-w}

\maketitle

\section{Introduction}

In the literature, there exists a plethora of non minimally
coupled gravity theories which are capable of producing viable
results. Such a theory is also the Einstein-Gauss-Bonnet theory
\cite{Hwang:2005hb,Nojiri:2006je,Cognola:2006sp,Nojiri:2005vv,Nojiri:2005jg,Satoh:2007gn,Bamba:2014zoa,Yi:2018gse,Guo:2009uk,Guo:2010jr,Jiang:2013gza,Kanti:2015pda,vandeBruck:2017voa,Kanti:1998jd,Pozdeeva:2020apf,Fomin:2020hfh,DeLaurentis:2015fea,Chervon:2019sey,Nozari:2017rta,Odintsov:2018zhw,Kawai:1998ab,Yi:2018dhl,vandeBruck:2016xvt,Kleihaus:2019rbg,Bakopoulos:2019tvc,Maeda:2011zn,Bakopoulos:2020dfg,Ai:2020peo,Odintsov:2019clh,Oikonomou:2020oil,Odintsov:2020xji,Oikonomou:2020sij,Odintsov:2020zkl,Odintsov:2020sqy,Odintsov:2020mkz,Easther:1996yd,Antoniadis:1993jc,Antoniadis:1990uu,Kanti:1995vq,Kanti:1997br},
once certain constraints are imposed. In order to achieve
viability, it is mandatory to equate the velocity of the
gravitational waves with the speed of light, as stated by the
GW170817 event \cite{GBM:2017lvd}. This case has been studied
thoroughly in
\cite{Odintsov:2019clh,Oikonomou:2020oil,Odintsov:2020xji,Odintsov:2020zkl,Odintsov:2020sqy,Oikonomou:2020sij,Odintsov:2020mkz}
assuming a slow-roll evolution. In the present paper, we shall
follow a physically different approach which, as it turns out, can
also lead to viable phenomenology. Assuming a constant-roll
evolution for the scalar field, which shall be implemented in the
present paper, the behavior of the scalar field $\phi$ is
completely different. Now, the rate of change of the scalar field
is decided by a single parameter, the constant-roll parameter.
This is also the decisive factor in the resulting values of the
observational quantities of inflation since for example the
spectral index of the primordial curvature perturbations is
strongly affected by this particular parameter, hence the reason
why the constant-roll assumption is convenient. In this paper,  we
shall develop and present the framework of constant-roll
Einstein-Gauss-Bonnet gravity in detail. After specifying the
gravitational action and deriving the equations of motion,
constraints on the velocity of the gravitational waves are imposed
in order to achieve a viable phenomenology compatible with the
GW170817. Combining this restriction and the constant-roll
assumption, a functional relation for scalar field is produced.
Then, the slow-varying parameters such as the slow-roll indices,
which play a significant role in our study, are introduced. In the
following sections, we present a variety of designations for the
scalar coupling functions following functional assumptions for the
equations of motion. In each case, the corresponding scalar
potential which is present in this framework, is derived directly
from the equations of motion and then we compare the results of
each model and study the behavior of the system under changes in
the values of the free parameters. Finally, we discuss and present
in detail all the possible approximations of the equations of
motion and in the appendix,  we include a pedagogical new
formalism which materializes various approximations in non-minimal
Einstein-Gauss-Bonnet gravity, and we make it available to the
interested reader.

At this point, it is also worth mentioning that in the present
paper, the cosmological background will be assumed to be that of a
flat Friedman-Robertson-Walker metric, with the corresponding line
element being,
\begin{equation}
\centering
\label{metric}
ds^2=-dt^2+\sum_{i=1}^{3}{(dx^i)^2},\,
\end{equation}
where $a(t)$ denotes the scale factor.

\section{Essential Features of Constant-Roll Inflation of non-minimally Coupled Einstein-Gauss-Bonnet Gravity}

In this section we shall present the formalism of inflationary
dynamics in the context of non-minimally coupled
Einstein-Gauss-Bonnet. The corresponding gravitational action is
assumed to be equal to,
\begin{equation}
\centering
\label{action}
S=\int{d^4x\sqrt{-g}\left(\frac{h(\phi)R}{2\kappa^2}-\frac{1}{2}\omega g^{\mu\nu}\partial_\mu\phi\partial_\nu\phi-V(\phi)-\xi(\phi)\mathcal{G}\right)},\,
\end{equation}
where $g$ is the determinant of the metric tensor $g_{\mu\nu}$
specified previously in the line element, $R$ denotes the Ricci
scalar, $h(\phi)$ is a dimensionless scalar function coupled to
the Ricci scalar and $V(\phi)$ signifies the scalar potential of
the scalar field and finally, the term $\xi(\phi)\mathcal{G}$
refers to the string corrections with $\mathcal{G}$ being the
Gauss-Bonnet invariant and $\xi(\phi)$ the Gauss-Bonnet coupling
scalar function. In particular, the Gauss-Bonnet invariant is
defined as
$\mathcal{G}=R^2-4R_{\alpha\beta}R^{\alpha\beta}+R_{\alpha\beta\gamma\delta}R^{\alpha\beta\gamma\delta}$
with $R_{\alpha\beta}$ and $R_{\alpha\beta\gamma\delta}$ being the
Ricci and Riemann tensor respectively and due to the form of the
line element, they are simplified to the expressions
$\mathcal{G}=24H^2(\dot H+H^2)$ and $R=12H^2+6\dot H$ where $H$
denotes Hubble's parameter defined as $H=\frac{\dot a}{a}$ and the
``dot'' as usual implies differentiation with respect to the
cosmic time. Furthermore, by assuming the scalar field $\phi$ is
homogeneous, meaning only time dependent, then the kinetic term is
also simplified to the form $-\frac{1}{2}\omega\dot\phi^2$.
Finally, we mention that in the following results, we shall assume
a canonical kinetic term which in turn implies that $\omega=1$,
but we shall keep it as it is in order to have the phantom or
non-canonical (but nevertheless constant) case available for the
reader.

Implementing the variation principle with respect to the metric
tensor and the scalar field in Eq. (\ref{action}) generates the
field equations of gravity and the continuity equation of the
scalar field. By splitting the field equations in time and space
components, the gravitational equations of motion are then derived
which read,
\begin{equation}
\centering
\label{motion1}
\frac{3hH^2}{\kappa^2}=\frac{1}{2}\omega\dot\phi^2+V-\frac{3H\dot h}{\kappa^2}+24\dot\xi H^3,\,
\end{equation}
\begin{equation}
\centering
\label{motion2}
-\frac{2h\dot H}{\kappa^2}=\omega\dot\phi^2-\frac{H\dot h}{\kappa^2}-16\dot\xi H\dot H+\frac{h''\dot\phi^2+h'\ddot\phi}{\kappa^2}-8H^2(\ddot\xi-H\dot\xi),\,
\end{equation}
\begin{equation}
\centering
\label{motion3}
\omega(\ddot\phi+3H\dot\phi)+V'-\frac{Rh'}{2\kappa^2}+24\xi'H^2(\dot H+H^2)=0,\,
\end{equation}
where with ``prime''at this instance, we denote differentiation
with respect to the scalar field. This system of equations
contains all the available information of the inflationary era,
but unfortunately it is very intricate and cannot be solved
analytically. One may proceed only by making certain
approximations which must hold true in order for the model to be
rendered viable. Furthermore, in order to be called viable, the
model obviously must not be at variance with the observations.
Recent observations such as the striking GW170817 event have made
it abundantly clear that the tensor  perturbations on the metric
propagate with a velocity equal to that of light's, and therefore
each model of modified gravity stating otherwise must be
discarded, no matter the mathematical beauty it may involve. The
modified Einstein-Gauss-Bonnet gravity is very promising since
even though it generates a formula for the velocity of the
gravitational waves which at first site seems to deviate from the
speed of light, but it can be carefully modified in order to
``survive'' the test of GW170817. In particular, the expression
for the velocity in natural units, where $c=1$, is,
\begin{equation}
\centering
\label{velocity}
c_T^2=1-\frac{Q_f}{2Q_t},\,
\end{equation}
where $Q_f$ and $Q_t$ are auxiliary functions defined as
$Q_f=16(\ddot\xi-H\dot\xi)$ and $Q_t=\frac{h}{\kappa^2}-8\dot\xi
H$. Thus, the theory can predict velocity for the gravitational
waves equal to the speed of light by equating the numerator of the
second term to zero, or in other words $\ddot\xi=H\dot\xi$. This
relation also simplifies the second equation of motion
(\ref{motion2}). Upon solving this ordinary differential equation,
one finds the expression for the derivative of the Gauss-Bonnet
coupling function. This method has been studied previously in
. However, there exists a different method which seems
to reveal deeper connections between the functions of the scalar
field. Since the differential operator with respect to time can be
replaced with one depending on the scalar field, i.e.
$\frac{d}{dt}=\dot\phi\frac{d}{d\phi}$, then the differential
equations can be rewritten as,
\begin{equation}
\centering
\xi''\dot\phi^2+\xi'\ddot\phi=H\xi'\dot\phi.\,
\end{equation}
From this equation, a deeper connection can be extracted by two
different assumptions. Assuming that the slow-roll approximation
holds, then $\ddot\phi\ll H\dot\phi$, thus the term
$\xi'\ddot\phi$ is negligible. This method has been studied
thoroughly both in the minimally and non-minimally coupled case,
see \cite{Odintsov:2020sqy} and \cite{Odintsov:2020xji}
respectively. An alternative approach is to assume a constant-roll
condition, which is also the one we shall implement. Letting
$\ddot\phi=\beta H\dot\phi$ where $\beta$ is the constant-roll
parameter, then the previous equation can be solved exactly. The
resulting expression is shown below,
\begin{equation}
\centering
\label{dotphi}
\dot\phi=H(1-\beta)\frac{\xi'}{\xi''},\,
\end{equation}
where it was obviously assumed that $\dot\phi$ is nonzero. This in
turn implies that $\beta\neq1$ and as we shall see, this
prohibition states the slow-roll indices do not drop to zero
simultaneously, however their values could be small. In addition,
in order to simplify the equations of motion, we shall assume that
the slow-roll approximations hold true. However, the value of
$\beta$ will not be assumed by definition to be small and will be
examined during the fine tuning of a model. Hence, we demand that,
\begin{align}
\centering
\label{slowrollapprox}
\dot H&\ll H^2,&\frac{1}{2}\omega\dot\phi&\ll V,\,
\end{align}
during the first horizon crossing, or the initial stage of
inflation. This assumption simplifies the equations of motion,
which in turn are rewritten as follows,
\begin{equation}
\centering
\label{motion4a}
\frac{3hH^2}{\kappa^2}\simeq V-\frac{3Hh'\dot\phi}{\kappa^2}+24\dot\xi H^3,\,
\end{equation}
\begin{equation}
\centering
\label{motion5a}
-\frac{2h\dot H}{\kappa^2}=\omega\dot\phi^2-\frac{Hh'\dot\phi}{\kappa^2}-16\xi'\dot\phi H\dot H+\frac{h''\dot\phi^2+\beta Hh'\dot\phi}{\kappa^2},\,
\end{equation}
\begin{equation}
\centering
\label{motion6a}
\omega(3+\beta)H\dot\phi+V'-6H^2\frac{h'}{\kappa^2}+24\xi'H^4\simeq0.\,
\end{equation}
However, even with the slow-roll approximations holding true, the
system of differential equations still remains intricate and
cannot be solved. Further approximations are needed in order to
derive the inflationary phenomenology, so by following Ref.
\cite{Odintsov:2020xji}, the equations of motion can be written as
follows,
\begin{equation}
\centering
\label{motion4}
\frac{3hH^2}{\kappa^2}\simeq V-\frac{3H^2(1-\beta)h'}{\kappa^2}\frac{\xi'}{\xi''},\,
\end{equation}
\begin{equation}
\centering
\label{motion5}
-\frac{2h\dot H}{\kappa^2}\simeq(1+\beta)^2H^2\frac{\xi'}{\xi''}\left(-\frac{h'}{\kappa^2}+(\omega+\frac{h''}{\kappa^2})\frac{\xi'}{\xi''}\right),\,
\end{equation}
\begin{equation}
\centering
\label{motion6}
V'+3H^2\left(\omega(1-\beta)(1+\frac{\beta}{3})\frac{\xi'}{\xi''}-2\frac{h'}{\kappa^2}\right)\simeq0.\,
\end{equation}
These are the simplified equations of motion which shall be
studied in the present paper. The benefit of this particular
approach is that equation (\ref{motion5}) is written
proportionally to the squared Hubble's parameter so, as we shall
explain in the following, it is the explicit form of slow-roll
index $\epsilon_1$. Furthermore, equation (\ref{motion6})
describes a differential equation of the scalar potential or the
Ricci coupling. This insinuates that the scalar functions, which
in the present paper are the two coupling functions $h(\phi)$ and
$\xi(\phi)$ and the scalar potential $V(\phi)$ as well, are
interconnected. Hence, we cannot designate freely each scalar
function but rather, by specifying two of them, the other can be
produced directly from the differential equation. In this paper,
we shall assume that the scalar potential is derivable from
Eq.(\ref{motion6}) while the coupling functions are designated
freely. Working differently however is also a possibility. Lastly,
we note that we shall study certain different approaches one can
follow in order to study the viability of a model under the
constant-roll condition and summarize the results. Let us now
proceed with the dynamics of inflation.

Previously, we referred to the slow-roll indices. These indices
are a powerful tool which helps us ascertain the validity of a
specific model. Since string corrections are implemented, the
degrees of freedom have been raised by two, hence the total number
of indices which must be evaluated are six as presented below,
\begin{align}
\centering
\epsilon_1&=\pm\frac{\dot H}{H^2},&\epsilon_2&=\frac{\ddot\phi}{H\dot\phi},&\epsilon_3&=\frac{\dot F}{2HF}, &\epsilon_4&=\frac{\dot E}{2HE},& \epsilon_5&=\frac{\dot F+Q_a}{2HQ_t},&\epsilon_6&=\frac{\dot Q_t}{2HQ_t},\,
\end{align}
where,
\begin{equation}
\centering
F=\frac{h}{\kappa^2},\,
\end{equation}
\begin{equation}
\centering
E=F\left(\omega+\frac{3}{2Q_t}\left(\frac{\dot F+Q_a}{\dot\phi}\right)^2\right),\,
\end{equation}
\begin{equation}
\centering
Q_a=-8\dot\xi H^2,\,
\end{equation}
\begin{equation}
\centering
Q_t=F-8\dot\xi H.\,
\end{equation}
The sign of the slow-roll index $\epsilon_1$ seems arbitrary, due
to the fact that in different approaches we shall follow, it is
more preferable to use the positive sign over the negative or vice
versa. According to the previous relations, the slow-roll indices
can be written as,
\begin{equation}
\centering
\label{index1}
\epsilon_1\simeq\pm\frac{(1+\beta)^2}{2h}\frac{\xi'}{\xi''}\left(h'-(\kappa^2\omega+h'')\frac{\xi'}{\xi''}\right),\,
\end{equation}
\begin{equation}
\centering
\label{index2}
\epsilon_2=\beta,\,
\end{equation}
\begin{equation}
\centering
\label{index3}
\epsilon_3=\frac{1-\beta}{2}\frac{\xi'}{\xi''}\frac{h'}{h},\,
\end{equation}
\begin{equation}
\centering
\label{index4}
\epsilon_4=\frac{1-\beta}{2}\frac{\xi'}{\xi''}\frac{E'}{E},\,
\end{equation}
\begin{equation}
\centering
\label{index5}
\epsilon_5=\frac{1-\beta}{2Q_t}\frac{\xi'}{\xi''}\left(\frac{h'}{\kappa^2}-8\xi'H^2\right),\,
\end{equation}
\begin{equation}
\centering
\label{index6}
\epsilon_6=\frac{1-\beta}{2Q_t}\frac{\xi'}{\xi''}\left(\frac{h'}{\kappa^2}-8\xi'(\dot H+H^2)\right).\,
\end{equation}
Since the slow-roll conditions are assumed to hold true, then
$\epsilon_6\simeq\epsilon_5$. Nevertheless, in the following
models we shall extract its full form and ascertain whether the
previous statements hold true by examining the numerical values.
These equations coincide with the slow-roll case, that is in the
limit where $\beta\to0$, however this does not apply to index
$\epsilon_2$ as it can easily be inferred. Also, a quick glimpse
at index $\epsilon_1$, which plays an important role in producing
results as we shall see in the following, should convince the
reader as to why it was deemed suitable to designate freely the
scalar coupling functions and derive the form of the scalar
potential from the aforementioned differential equation in
(\ref{motion6}). It is also worth writing down the explicit forms
of the previously defined auxiliary functions,
\begin{equation}
\centering
Q_a=-8(1-\beta)\frac{\xi'^2}{\xi''}H^3,\,
\end{equation}
\begin{equation}
\centering
E=\frac{h\omega}{\kappa^2}+\frac{3h\xi''}{2\left(h\xi''-8(1-\beta)(\kappa\xi'H)^2\right)}\left(\frac{h'}{\kappa^2}-8\xi'H^2\right)^2,\,
\end{equation}
\begin{equation}
\centering
Q_t=\frac{h}{\kappa^2}-8(1-\beta)\frac{\xi'^2}{\xi''}H^2,\,
\end{equation}
\begin{equation}
\centering
Q_e=-32(1-\beta)\frac{\xi'^2}{\xi''}H\dot H.\,
\end{equation}
It is convenient to introduce the auxiliary function $Q_e$ as it
will appear in subsequent calculations. Finally, we discuss the
strategy we shall follow in order to study the viability of a
model by following a certain approach. Even if the equations of
motion have been simplified greatly, there are cases where further
simplifications lead to decent phenomenology and is also
justifiable. This can be inferred easily from Eq. (\ref{index1})
where neglecting certain terms leads to a different appearance of
the slow-roll index $\epsilon_1$ and consequently in completely
different phenomenology. This also applies to the form of Hubble's
parameter in (\ref{motion4}) and the differential equation of the
scalar potential (\ref{motion6}) as well. The main goal of our
study is to find the most productive approach one can follow in
order to find an appealing form of the slow-roll index
$\epsilon_1$. Subsequently, we evaluate the final value of the
scalar field during the inflationary era by assuming index
$\epsilon_1$ becomes of order $\mathcal{O}(1)$, or in other words
by equating $\epsilon_1=1$, but firstly we specify properly the
sign of the index which suits us better. Continuing, we shall
extract the value of the scalar field during the first horizon
crossing. This can be achieved easily by implementing the
e-folding number. By definition, the $e$-foldings number is equal
to $N=\int_{t_i}^{t_f}{Hdt}$ where the difference $t_f-t_i$
signifies the time duration of the inflationary era. By taking
advantage of the differential operator
$\frac{d}{dt}=\dot\phi\frac{d}{d\phi}$ and also, the form of
$\dot\phi$ in Eq. (\ref{dotphi}) the e-folding number is written
as,
\begin{equation}
\centering
\label{efolds}
N=\frac{1}{1-\beta}\int_{\phi_i}^{\phi_f}{\frac{\xi''}{\xi'}d\phi}.\,
\end{equation}
From this equation, the initial value of the scalar field can be
extracted easily. This is another reason as to why it is
convenient to define the Gauss-Bonnet coupling scalar function
instead of deriving it from a differential equation. Subsequently,
this value will be taken as an input in the observational indices
of inflation, which are the spectral index of primordial curvature
perturbations $n_S$, the spectral index of tensor perturbations
$n_T$ and the tensor-to-scalar ratio $r$, and these are defined
as,
\begin{align}
\centering
\label{results}
n_s&=1+2\frac{2\frac{\dot H}{H^2}-\epsilon_2+\epsilon_3-\epsilon_4}{1+\frac{\dot H}{H^2}},& n_T&=2\frac{\frac{\dot H}{H^2}-\epsilon_6}{1+\frac{\dot H}{H^2}},&r&=16\left|\left(\frac{Q_e}{4HF}+\frac{\dot H}{H^2}-\epsilon_3\right)\frac{Fc_A^3}{Q_t}\right|,\,
\end{align}
where $c_A$ is the sound wave speed defined as,
\begin{equation}
\centering
\label{soundwave1}
c_A^2=1+\frac{(\dot F+Q_a)Q_e}{3(\dot F+Q_a)^2+2\omega\dot\phi^2Q_t}.\,
\end{equation}
According to the latest Planck 2018 collaboration
\cite{Akrami:2018odb}, the measured values of these quantities
are,
\begin{align}
\centering
n_S&=0.9649\pm0.0042,&r&<0.064.\,
\end{align}
The spectral index of the tensor perturbations does not have a
value since B-modes have yet to be observed, but we shall also
derive the expected numerical value for each model. These are also
the values that we shall try and produce in each model, while
simultaneously respecting the necessary approximations which were
made. In addition, we shall also derive expressions for the amount
of non-Gaussianities in the primordial power spectrum.

\section{Inspecting Non Gaussianities  in view of constant-roll condition}

So far, the latest experiments have provided us with only limits
for the hypothesized non-Gaussianities. Until this point, it seems
that the curvature perturbations are described decently by a
Gaussian distribution, hence if existent, the amount of
non-Gaussianities must be quite small, hence with small deviation
from the Gaussian distribution. We can quantify the amount of the
expected non-Gaussianities by making use of additional auxiliary
parameters. We define \cite{DeFelice:2011zh}
\begin{align}
\centering
\label{nonGaussterms}
\delta_h&=\frac{\dot h}{Hh}=2\epsilon_3,
&\delta_\xi&=\frac{\kappa^2H\dot\xi}{h},
&\delta_X&=\frac{\kappa^2\omega\dot\phi^2}{2hH^2},
&\epsilon_s&=-\frac{\dot H}{H^2}+\frac{\delta_h}{2}-4\delta_\xi,
&\eta&=\frac{\dot\epsilon_s}{H\epsilon_s},
&s&=\frac{\dot c_A}{Hc_A}.\,
\end{align}
These parameters are obviously depending on Hubble's parameter and
its derivative. Hence, they are strongly depending on the approach
one shall follow and no certain expression can be produced.
However we can present their generalized forms depending on the
coupling scalar functions as shown below,
\begin{equation}
\centering
\delta_h=(1-\beta)\frac{\xi'}{\xi''}\frac{h'}{h},\,
\end{equation}
\begin{equation}
\centering
\delta_\xi=\frac{\kappa^2H^2(1-\beta)}{h}\frac{\xi'^2}{\xi''},\,
\end{equation}
\begin{equation}
\centering
\delta_X=\frac{\kappa^2\omega}{2}\frac{(1-\beta)^2}{h}\left(\frac{\xi'}{\xi''}\right)^2,\,
\end{equation}
\begin{equation}
\centering
\epsilon_s=-\frac{\dot H}{H^2}+\frac{1-\beta}{h}\frac{\xi'}{\xi''}\left(\frac{h'}{2}-\xi'(2\kappa H)^2\right),\,
\end{equation}
\begin{equation}
\centering
\eta=(1-\beta)\frac{\xi'}{\xi''}\frac{\epsilon_s'}{\epsilon_s},\,
\end{equation}
\begin{equation}
\centering
s=(1-\beta)\frac{\xi'}{\xi''}\frac{c_A'}{c_A},\,
\end{equation}
\begin{equation}
\centering
\label{soundwave2}
c_A^2\simeq1-\frac{2\delta_\xi(\delta_h-8\delta_\xi)(3\delta_h-24\delta_\xi-4\delta_X)}{\delta_X}.\,
\end{equation}
Here, we note that $c_A$ is again the sound wave velocity
specified differently with respect to the newly defined
parameters. However it is still equivalent to Eq.
(\ref{soundwave1}). From these forms, it can easily be inferred
that an appropriate designation of the Gauss-Bonnet coupled scalar
function which leads to a simple connection between its first two
derivatives simplifies the ratio and subsequently the forms of the
terms depending on such ratio. If accompanied by an appropriate
Ricci scalar coupling function, the resulting forms could be
functional and appealing. Let us now proceed with the definition
of the parameters which quantify the amount of non Gaussianities.
We define as the equilateral non-linear term $f_{NL}^{eq}$, the
spectra $\mathcal{P}_S$ the bispectrum  $\mathcal{B}_S$ of the
curvature perturbations and the three-point correlator
$\left<\mathcal{R}^3\right>$ as \cite{DeFelice:2011zh},
\begin{equation}
\centering
\label{NL}
f_{NL}^{eq}\simeq\frac{55}{36}\epsilon_s+\frac{5}{12}\eta+\frac{10}{3}\delta_\xi,\,
\end{equation}
\begin{equation}
\centering
\label{spectra}
\mathcal{P}_S=\frac{\kappa^2H^2}{8\pi^2h\epsilon_sc_A},\,
\end{equation}
\begin{equation}
\centering
\label{bispectrum}
\mathcal{B}_S=\frac{3}{10}f_{NL}^{eq}(2\pi)^4\mathcal{P}_S^2,\,
\end{equation}
\begin{equation}
\centering
\label{correlator}
\left<\mathcal{R}^3\right>=(2\pi)^3\mathcal{B}_S.\,
\end{equation}
These relations are the final product which we shall use. The
amount of non-Gaussianities shall be described by finding the
value of the equilateral non-linear term $f_{NL}^{eq}$ and the
three-point correlator $\left<\mathcal{R}^3\right>$. Typically,
these terms are evaluated in the Fourier space by inserting three
wavenumbers. Here we shall use the same wavenumber, hence the
equilateral part, which is obviously the wavenumber during the
first horizon crossing. Following the same steps as before, the
amount of non-Gaussianities can be specified by inserting the
value of the scalar field during the first horizon crossing,
exactly as in the case of the spectral indices and the
tensor-to-scalar ratio. Furthermore, we define a last parameter,
the skewness, which is indicative of the deviation of the
curvature perturbations from the Gaussian distribution and since
the bispectrum and the spectra are scale invariant, it can be
specified as,
\begin{equation}
\centering
\label{skewness}
\mathcal{S}\sim\mathcal{B}_S\sqrt{\mathcal{P}_S}.\,
\end{equation}
As a last comment, we mention that both spectral indices and the
tensor-to-scalar ratio can be derived from the new parameters
defined in (\ref{nonGaussterms}) as,
\begin{align}
\centering
n_S&=1+2\frac{\dot H}{H^2}-\delta_h-\eta-s,
&n_T&=-2\epsilon_s-8\delta_\xi,
&r&=\frac{16\epsilon_sc_A}{1-8\delta_\xi},\,
\end{align}
which are exactly equivalent to the previous definition
(\ref{results}) with respect to the slow-roll indices which is
also the one which we shall use as well. In the following we shall
examine the viability of certain models under specific approaches
in the equations and moreover predict the amount of
non-Gaussianities in each model separately.

\section{Testing the Compatibility of a model and predicting non-Gaussianities}

In the next subsections, we shall examine the viability of certain
models by comparing their produced results with the latest
observations and also make predictions about the existence of
non-Gaussianities. The possible approaches one can follow are many
but we shall present all of the available forms of Hubble's
parameter and its derivative, in at least three models. The
strategy is to designate the coupling functions and also choose
the form of Hubble's parameter and its derivative. Afterwards, we
shall derive the scalar potential from the differential equation,
\begin{equation}
\centering
\label{Vdif}
V'+3H^2\left(\omega(1-\beta)(1+\frac{\beta}{3})\frac{\xi'}{\xi''}-2\frac{h'}{\kappa^2}\right)\simeq0,\,
\end{equation}
and upon evaluating the spectral indices, the tensor-to-scalar
ratio and predict the amount of non-Gaussianities from the value
of the scalar field during the first horizon crossing, we shall
ascertain whether the approximations made apply and in consequence
whether the model is rendered viable. In certain cases we shall
use even a simpler differential equation for the scalar potential
where one omits the ratio $\xi'/\xi''$. This in turn results in a
simpler expressions for the potential. Also, when feasible, we
shall make comparisons between the slow-roll and constant-roll
condition.

\subsection{Model I: Linear and Exponential Couplings}

Suppose that the coupling scalar functions are defined as,
\begin{equation}
\centering
\label{h1}
h(\phi)=\Lambda\kappa\phi,\,
\end{equation}
\begin{equation}
\centering
\label{xi1}
\xi(\phi)=\lambda e^{\gamma\kappa\phi}.\,
\end{equation}
Here, each parameter is dimensionless. These functions were chosen
due to the appealing characteristics they poses. The first
function has a zeroth second derivative, which simplifies greatly
Eq. (\ref{index1}) and also, the ratio $\xi'/\xi''$ appears to be
constant as,
\begin{equation}
\centering
\xi''=\gamma\kappa\xi'.\,
\end{equation}
As a result, all the previous equations will indeed simplify. At
this point, it is worth mentioning the approach we shall choose.
For Hubble's parameter and its derivative, we shall choose the
following forms,
\begin{equation}
\centering
\label{motion1A}
H^2\simeq\frac{\kappa^2V}{3h\left(1+(1-\beta)\frac{h'}{h}\frac{\xi'}{\xi''}\right)},\,
\end{equation}
\begin{equation}
\centering
\label{motion2A}
\dot H\simeq\frac{H^2}{2}(1-\beta)^2\frac{\xi'}{\xi''}\left(\frac{h'}{h}-\frac{\kappa^2\omega}{h}\frac{\xi'}{\xi''}\right).\,
\end{equation}
This case can in fact be though of as the complete case where the
term $h''$ is naturally equal to zero due to the linear choice of
the Ricci coupling function. Moreover, it is convenient to choose
the positive sign for index $\epsilon_1$ in this case due to the
form of Hubble's derivative. Concerning the scalar potential, from
Eq. (\ref{Vdif}) it can be inferred that,
\begin{equation}
\centering
\label{potA}
V(\phi)=V_0(\beta -\gamma  \kappa  \phi -1)^{\alpha_1},\,
\end{equation}
where $\alpha_1=\frac{\left(\beta ^2+2 \beta -3\right) \omega +6
\gamma  \Lambda }{3 \gamma  \Lambda } $ and $V_0$ is the
integration constant with mass dimensions $[m]^{4}$. This form can
be categorized as a power-law model as well with an exponent which
may not be integer necessarily as expected. Let us proceed with
the evaluation of certain auxiliary parameters and the slow-roll
indices. According to their definitions, we get,
\begin{equation}
\centering
\label{dXA}
\delta_X=\frac{\omega(1-\beta)^2}{2\gamma h(\phi)},\,
\end{equation}
\begin{equation}
\centering
\label{dxiA}
\delta_\xi\simeq\frac{(1-\beta)\xi(\phi)\kappa^4V(\phi)}{3h^2(\phi)\left(1+\frac{1-\beta}{\gamma\kappa\phi}\right)},\,
\end{equation}
\begin{equation}
\centering
\label{esA}
\epsilon_s\simeq-\frac{\Lambda(1-\beta)^2}{2\gamma h(\phi)}\left(1-\frac{\omega}{\Lambda\gamma}\right)+\frac{1-\beta}{\gamma h(\phi)}\left(\frac{\Lambda}{2h(\phi)}-\frac{4\gamma\xi(\phi)\kappa^4V(\phi)}{3\left(h(\phi)+\frac{(1-\beta)\Lambda}{\gamma}\right)}\right),\,
\end{equation}
\begin{equation}
\centering
\label{index1A}
\epsilon_1\simeq\frac{(1-\beta )^2 (\gamma  \Lambda -\omega )}{2 \gamma ^2 \kappa  \Lambda  \phi },\,
\end{equation}
\begin{equation}
\centering
\label{index2A}
\epsilon_2=\beta,\,
\end{equation}
\begin{equation}
\centering
\label{index3A}
\epsilon_3=\frac{1-\beta}{2 \gamma  \kappa  \phi },\,
\end{equation}
\begin{equation}
\centering
\label{index5A}
\epsilon_5=\frac{(1-\beta ) \left(8 \gamma ^2 \lambda  \kappa^4V(\phi ) e^{\gamma  \kappa  \phi }-3 \Lambda ^2 (1-\beta +\gamma  \kappa  \phi )\right)}{2 \gamma    \left( 8 (1-\beta ) \gamma  \lambda \kappa^4  V(\phi ) e^{\gamma  \kappa  \phi }-3 \Lambda ^2 \kappa\phi  (1-\beta +\gamma  \kappa  \phi )\right)},\,
\end{equation}
\begin{equation}
\centering
\label{index6A}
\epsilon_6=\frac{(1-\beta ) \left(3 \Lambda ^2 (-\beta +\gamma  \kappa  \phi +1)^2-8 (1-\beta ) \gamma  \lambda  e^{\gamma  \kappa  \phi } \left( \kappa ^3V'(\phi )-(\beta -\gamma  \kappa  \phi ) \left( \kappa ^3V'(\phi )+\gamma  \kappa^4  V(\phi )\right)\right)\right)}{2 \gamma  (1-\beta +\gamma  \kappa  \phi ) \left(3 \Lambda ^2 \kappa\phi  (1-\beta +\gamma  \kappa  \phi )-8 (1-\beta ) \gamma   \lambda  \kappa ^4 V(\phi ) e^{\gamma  \kappa  \phi }\right)}.\,
\end{equation}
Only these parameters can be written in explicit forms depending
on the scalar field $\phi$ and in certain cases, $\eta$ as well.
The rest have very intricate forms and there is no use in writing
them analytically. As written, it is obvious that only the first
three slow-roll indices have functional forms compared to the rest
parameters. Due to the functionality of $\epsilon_1$ however, the
initial and final value of the scalar field can be extracted
easily so there is no need to worry about the rest slow-roll
indices. Hence, by equating $\epsilon_1$ to unity and by making
use of Eq. (\ref{efolds}), we end up with the following two
expressions,
\begin{equation}
\centering
\label{scalarfA}
\phi_f=\frac{(1-\beta )^2 (\gamma  \Lambda -\omega )}{2 \gamma ^2 \kappa  \Lambda },\,
\end{equation}
\begin{equation}
\centering
\label{scalariA}
\phi_i=\frac{(1-\beta ) (\gamma  \Lambda  (1-\beta -2 N)-(1-\beta)\omega )}{2 \gamma ^2 \kappa  \Lambda }.\,
\end{equation}
The final formula is also the one which as stated before, shall be
used as an input in Eq. (\ref{results}), Assuming that in Planck
Units, i.e $\kappa^2=1$, the free parameters of the theory obtain
the value ($\omega$, $\lambda$, $\Lambda$, $V_0$, $N$, $\gamma$,
$\beta$)=(1, 10$^{-4}$, 100, $10^{-6}$, 60, -100, -0.008) then the
resulting values for the spectral indices and the tensor-to-scalar
ratio are $n_S=0.965075$, $n_T=-0.0001373$ and $r=0.001089$ which
are obviously compatible with recent Planck observational data
\cite{Akrami:2018odb}. Concerning the scalar field itself, we
mention that in Planck Units, $\phi_i=0.599719$ and
$\phi_f=-0.00508$ which indicates a decreasing with time
homogeneous scalar field. Moreover, for the numerical values of
the slow-roll indices, we mention that $\epsilon_1=-0.008472$,
$\epsilon_3=-0.008404$, $\epsilon_5=-0.000033$,
$\epsilon_5=-0.008404=\epsilon_6$, which in turns shows that the
slow-roll approximations hold.

In addition to the acceptable value for the spectral index of
primordial curvature perturbations, we mention that the amount of
non-Gaussianities of the curvature perturbations due to this
particular choice of parameters is predicted to be
$f_{NL}^{eq}=0.00710729$ and
$\left<\mathcal{R}^3\right>=3.06672\cdot10^{-6}$ which as expected
are extremely small values. In fact, the corresponding deviation
from the Gaussian distribution, since the spectra $\mathcal{P}_S$
and the bispectrum $\mathcal{B}_S$ are scale independent, is
approximately $\mathcal{S}\sim9.65\cdot 10^{-11}$. Additionally,
$\delta_\xi=2.976\cdot10^{-37}$, $\delta_X=1.6656\cdot10^{-11}$,
$\eta=0.0168079$, $s=1.57743\cdot10^{-29}$ and
$\epsilon_s=0.000068$ which showcases that these auxiliary
parameters have extremely small values, apart from only two. These
two dominant parameters are also the ones who mainly define the
value of the non-linear term and subsequently, the three-point
correlator.
\begin{figure}[h!]
\centering
\includegraphics[width=20pc]{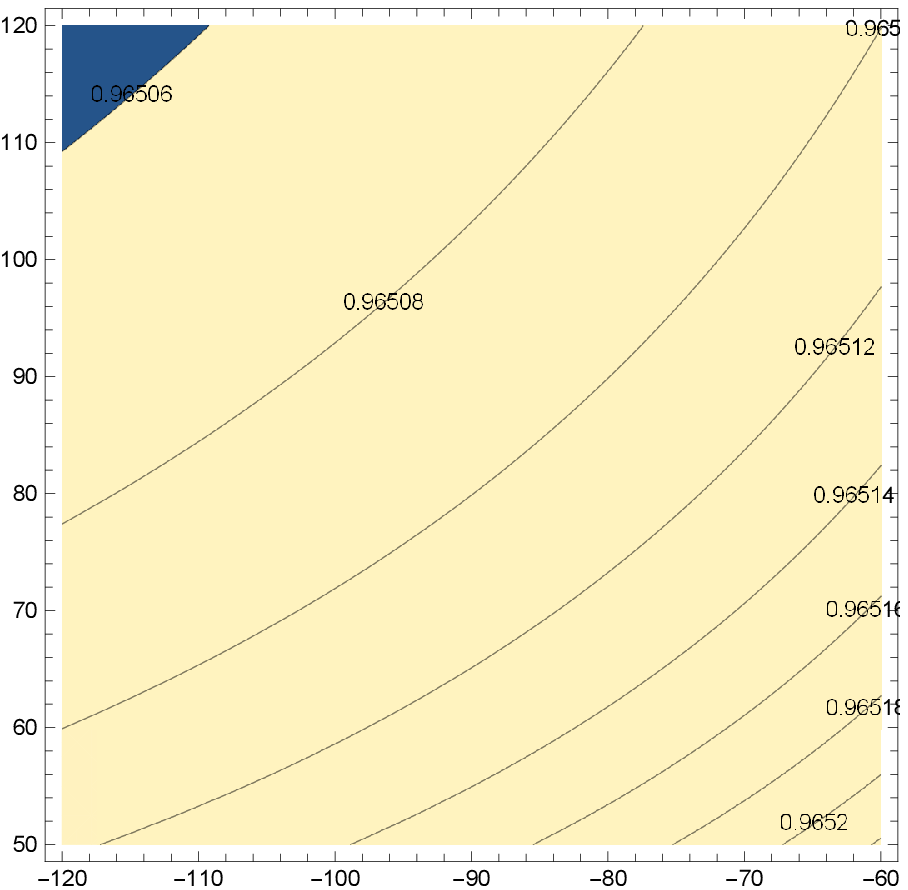}
\includegraphics[width=20pc]{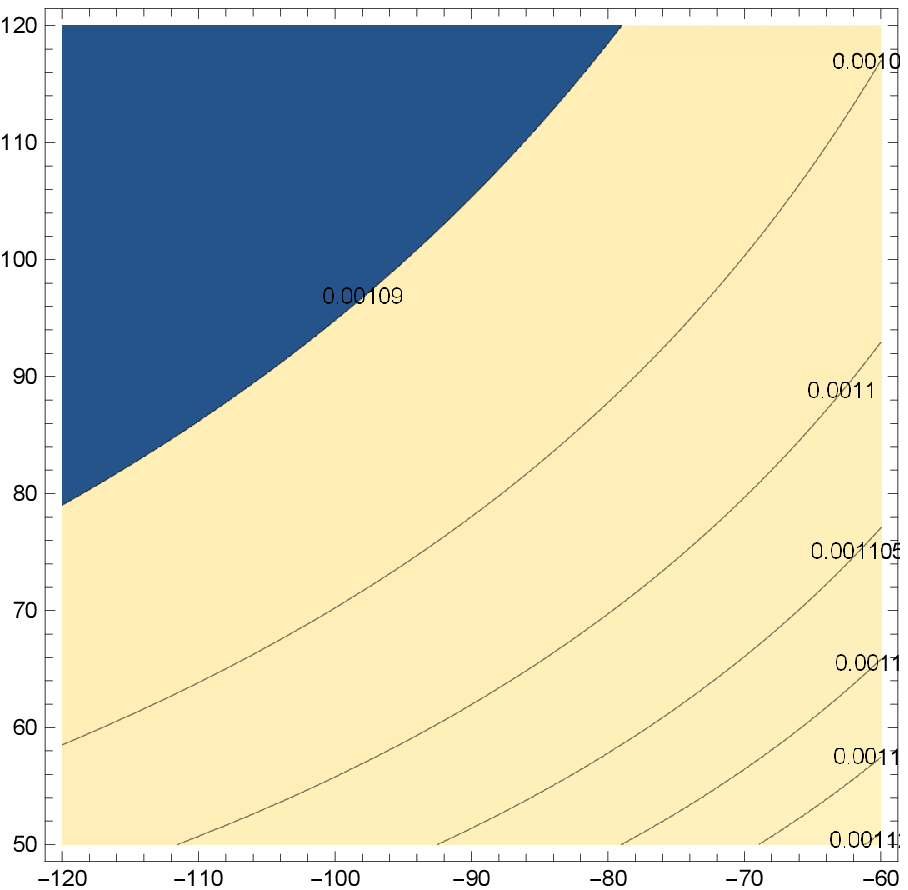}
\caption{Contour plots of the spectral index of scalar perturbations $n_S$ (left) and the tensor-to-scalar ratio $r$ (right) depending on parameters $\gamma$ and $\Lambda$. Their values range from [-120,-60] and [50,120] respectively. These plots indicate that even though a dependence is observed, the numerical value changes by a fraction in each case.}
\label{plot1}
\end{figure}
Examining the correspondence of the system under a change in the
values of the free parameters, once finds that the constant-roll
parameter $\beta$ has the most impact on the results while
$\Lambda$ and $\gamma$ influence them as well but not at a
significant rate as the first. For instance, we mention that
changing $\beta$ from -0.008 to -0.005, the spectral index of
primordial curvature perturbations obtains the value
$n_S=0.959128$ which makes this value incompatible with the
observations. We also note that the tensor-to-scalar ratio
experiences a change as well, but it decreases its value so it
remains acceptable. On the other hand, $\Lambda$ changes the
results but only when its order of magnitude alters and not its
sign. Numerically speaking, the value $\Lambda=-100$ produces the
results $n_S=0.964947$ and $r=0.00106215$ while $\Lambda=1$ gives
$n_S=0.969504$ and $r=0.0024329$, which is a more significant
change than the one produced by flipping the sign. The same
applies to  $\gamma$ as $\gamma=100$ gives the values
$n_S=0.964947$, $r=0.00106215$ while $\gamma=-1$ the values
$n_S=0.969504$ and $r=0.00243129$. This in turn implies that
$\Lambda$ and $\gamma$ have a similar behavior as the same changes
lead to the exact same numerical values. However, the most
dominant parameter which is also the decisive factor remains the
constant-roll parameter. The phenomenology of the model is
depicted in Fig. \ref{plot1} for some ranges of values of the free
parameters.

In a similar manner, the same parameters define the amount of
non-Gaussianities as well. This time however, the effects are
reversed. Each change in the parameters, as before, leads only to
an insignificant change in the fourth decimal (or even lesser) of
the value of the non linear term $f_{NL}^{eq}$. Furthermore,
constant-roll parameter $\beta$ has the same result in the
three-point correlator. However, $\Lambda$ and $\gamma$ now lead
to significant changes in the three-point correlator, since
changing $\Lambda=1$ leads to the value
$\left<\mathcal{R}^2\right>=67.967$ and changing $\gamma=-1$ leads
to $\left<\mathcal{R}^2\right>=6.7967\cdot 10^{-15}$. Thus, the
order of magnitude of the three-point correlator depends greatly
on $\Lambda$, $\gamma$ and in fact $V_0$, since the correlator is
proportional to squared Hubble's parameter. This is also the
reason why the constant in Ricci's coupling function affects its
order of magnitude as well. For example, a decrease in the
amplitude of the scalar potential by two orders leads to a
decrease in the order of magnitude of the three point correlator
by four orders.

Due to this particular choice of parameters, the exponent of the
scalar potential in Eq. (\ref{potA}) is 2.0001, which is really
close to 2. Hence, this particular model can be thought of as a
linear Ricci coupling, squared power-law potential and exponential
Gauss-Bonnet coupling function, which in principle is a simple
case. In other words, the simplest case in the choice of coupling
functions leads to a simple scalar potential as well. However,
this is not a universal characteristic, but rather a model
dependent one. If one were to use a different approach for
Hubble's parameter then the resulting scalar potential may not
have been that simple.

Finally, we examine the validity of the approximations which were
made necessarily in order to solve approximately the system of
equations of motion. Referring to the slow-roll approximations
which were assumed, we mention that during the first horizon
crossing, $\dot H/H^2\sim\mathcal{O}(10^{-3})$,
$\frac{1}{2}\omega\dot\phi^2\sim\mathcal{O}(10^{-10})$ and
$V\sim\mathcal{O}(10^{-3})$ which indicates that indeed the
slow-roll approximations do apply. However, these were not the
only approximations assumed in the equations of motion. Recalling
our previous statements, we mentioned that the string corrective
terms in each separate equation of motion will be omitted. This
again is a valid assumption, as it turns out that $24\dot\xi
H^3\sim\mathcal{O}(10^{-39})$, $16\dot\xi H\dot
H\sim\mathcal{O}(10^{-41})$  and
$24\xi'H^4\sim\mathcal{O}(10^{-37})$ which justifies why, compared
to the rest terms in Equations (\ref{motion1}), (\ref{motion2})
and (\ref{motion3}) respectively were omitted.

\subsection{Model II: Trigonometric and Power Law Couplings}

Suppose now that the scalar coupling functions are defined as,
\begin{equation}
\centering
\label{h2}
h(\phi)=\Lambda \sin(\gamma\kappa\phi+\theta),\,
\end{equation}
\begin{equation}
\centering
\label{xi2}
\xi(\phi)=\lambda(\kappa\phi)^m.\,
\end{equation}
This model was studied also in the slow-roll case in Ref.
\cite{Odintsov:2020xji} and was deemed as a false positive model,
as it generated compatible results while simultaneously violating
the approximations made in the equations of motion. This is a
bizarre choice for the Ricci coupling constant so it was tempting
to study the viability of the same model under the constant-roll
condition. For the equations of motion, we choose,
\begin{equation}
\centering
\label{motion1B}
H^2\simeq\frac{\kappa^2V}{3(1-\beta)h'}\frac{\xi''}{\xi'},\,
\end{equation}
\begin{equation}
\centering
\label{motion2B}
\dot H\simeq-\frac{H^2}{2}(1-\beta)^2\frac{h''}{h}\left(\frac{\xi'}{\xi''}\right)^2,\,
\end{equation}
\begin{equation}
\centering
\label{motion3B}
V'+3H^2\left(\omega(1-\beta)(1+\frac{\beta}{3})\frac{\xi'}{\xi''}-2\frac{h'}{\kappa^2}\right)\simeq0.\,
\end{equation}
It is clear that the trigonometric choice is perfect for such
approach since is simplifies greatly the ratio of the Ricci
coupling and in addition, since its second derivative is negative,
it is convenient to choose the positive sign of index
$\epsilon_1$. From this choice of coupling constant, it turns out
that the scalar potential is,
\begin{equation}
\centering
\label{potB}
V(\phi)=V_0(\kappa\phi)^{\alpha_1}\left(\frac{\cos(\frac{\gamma\kappa\phi+\theta}{2})-\sin(\frac{\gamma\kappa\phi+\theta}{2})}{\cos(\frac{\gamma\kappa\phi+\theta}{2})+\sin(\frac{\gamma\kappa\phi+\theta}{2})}\right)^{\alpha_2},\,
\end{equation}
where $\alpha_1=\frac{2(m-1)}{1-\beta}$ and
$\alpha_2=\frac{\omega(3+\beta)}{3\Lambda\gamma^2}$. It is the
same form as in the slow-roll condition where now the
constant-roll parameter $\beta$ appears in the exponents. The same
obviously applies to the rest parameters as well,
\begin{equation}
\centering
\label{dXB}
\delta_X=\frac{\omega}{2h(\phi)}\left(\frac{(1-\beta)\kappa\phi}{m-1}\right)^2,\,
\end{equation}
\begin{equation}
\centering
\label{dxiB}
\delta_\xi\simeq m(1-\beta)\frac{\xi(\phi)}{h(\phi)}\frac{\kappa^4V(\phi)}{3\gamma\kappa\phi\Lambda\cos(\gamma\kappa\phi+\theta)},\,
\end{equation}
\begin{equation}
\centering
\label{esB}
\epsilon_s\simeq\left(\frac{(1-\beta)\gamma\kappa\phi}{m-1}\right)^2+\frac{1-\beta}{(m-1)h(\phi)}\left(\frac{\Lambda\gamma\kappa\phi\cos(\gamma\kappa\phi+\theta)}{2}-\frac{4m\xi(\phi)\kappa^4V(\phi)}{3h(\phi)}\right),\,
\end{equation}
\begin{equation}
\centering
\label{index1B}
\epsilon_1\simeq\frac{1}{2}\left(\frac{(1-\beta)\gamma\kappa\phi }{ m-1}\right)^2,\,
\end{equation}
\begin{equation}
\centering
\label{index2B}
\epsilon_2=\beta,\,
\end{equation}
\begin{equation}
\centering
\label{index3B}
\epsilon_3=\frac{(1-\beta ) \gamma  \kappa  \phi  \cot (\gamma  \kappa  \phi +\theta )}{2 (m-1)},\,
\end{equation}
\begin{equation}
\centering
\label{index5B}
\epsilon_5=\frac{3 (\beta -1) \gamma ^2 \kappa  \Lambda ^2 \phi ^2 \cos ^2(\gamma  \kappa  \phi +\theta )+8 \kappa ^3 \lambda  (m-1) m V(\phi ) (\kappa  \phi )^m}{(m-1) \left(16 \kappa ^3 \lambda  m V(\phi ) (\kappa  \phi )^m-3 \gamma  \Lambda ^2 \phi  \sin (2 (\gamma  \kappa  \phi +\theta ))\right)},\,
\end{equation}
\begin{equation}
\centering
\label{index6B}
\epsilon_6\simeq\epsilon_5.\,
\end{equation}
In this case, even $\epsilon_6$ has a gory form so invoking the
slow-roll approximations, this index must be equal to
$\epsilon_5$. Finally we note that the values of the scalar field
at the initial and final stage of inflation are,
\begin{equation}
\centering
\label{scalarfB}
\phi_f=\pm\frac{\sqrt{2}}{\kappa}\left|\frac{m-1}{\gamma(1-\beta)}\right|,\,
\end{equation}
\begin{equation}
\centering
\label{scalariB}
\phi_i=\phi_fe^{-\frac{N(1-\beta)}{m-1}}.\,
\end{equation}
Here, we shall implement the positive value of $\phi_f$ and in
consequence, $\phi_i$. Letting ($\omega$, $\Lambda$, $\lambda$,
$V_0$, $N$, $m$, $\gamma$, $\theta$, $\beta$)=(1, 0.1, $10^{-14}$,
$10^{33}$, 60, 6, 2.5, $\frac{\pi}{6}$, 0.0174) then the produced
results are $n_S=0.965218$, $n_T=-0.0000185$ and $r=0.0001483$
which are compatible with the latest observations. In addition,
the predicted values for the non-Gaussianities are
$f_{NL}^{eq}=0.0818867$ and
$\left<\mathcal{R}^3\right>=2.03855\cdot10^{-6}$ with a
corresponding deviation $\mathcal{S}\sim3.14567\cdot10^{-11}$ from
the Gaussian distribution. The scalar field now seems to increase
with time since $\phi_i=0.00002179$ and $\phi_f=2.87851$ and
furthermore, the slow-roll criteria apply more than enough since
$\epsilon_1=5.73184\cdot10^{-11}$,
$\epsilon_3=9.27124\cdot10^{-6}$,
$\epsilon_4=2.40493\cdot10^{-7}$,
$\epsilon_5=9.27124\cdot10^{-6}=\epsilon_6$. Moreover
$\delta_\xi=1.35\cdot10^{-50}$, $\delta_X=9.8356\cdot10^{-20}$,
$\epsilon_s=9.27118\cdot10^{-6}$, $s=-1.67\cdot10^{-40}$ and
$\eta=0.196494$ which goes to say that only one parameter is
dominant.

However, even in the constant-roll case, the model is in fact
false since certain approximations are invalid, We note that the
slow-roll conditions do apply, however in order to derive equation
(\ref{motion1B}) from (\ref{motion1}), we made the assumption that
the term $Hh'$ is more dominant than the term $H^2h$, a reasonable
assumption which is unfortunately invalid as
$Hh'\sim\mathcal{O}(10^{-15})$ while
$H^2h\sim\mathcal{O}(10^{-11})$, hence the latter is more dominant
and cannot be discarded. Nevertheless, even if a single
approximation does not apply, the model is automatically rendered
invalid, no matter the beauty or functionality it may contain.
Perhaps a different set of values for the free parameters could
save this model in the end.

It is intriguing however to examine whether using a different form
for Hubble's parameter, mainly the condition which was invalid,
could manifest viable results. We shall also make a different
assumption as in the previous, where we neglect the ratio
$\xi'/\xi''$ in the differential equation of the scalar potential.
Suppose now that,
\begin{equation}
\centering
\label{motion1B2}
H^2\simeq\frac{\kappa^2V}{3h},\,
\end{equation}
\begin{equation}
\centering
\label{motion3B2}
V'-6H^2\frac{h'}{\kappa^2}\simeq0.\,
\end{equation}
Subsequently, the scalar potential reads,
\begin{equation}
\centering
\label{potB2}
V(\phi)=V_0\sin(\gamma\kappa\phi+\theta)^2,\,
\end{equation}
where for consistency, the integration constant $V_0$ has mass
dimensions [m]$^{4}$. A quick comparison between the two
potentials indicates that in the second approach, the potential is
functionally more elegant. This can be attributed to the  change
in the differential equation of the scalar field and not the
change in Hubble's parameter. Furthermore, changing Hubble and
also keeping the ratio $\xi'/\xi''$ in the differential equation
results in a perplexed potential similar to the previous case.
Hence, this simplification in the scalar potential is a universal
feat of this particular formalism we decided to work and is not
model dependent.

Concerning the auxiliary parameters and the slow-roll indices, we
mention that since they depend strongly on the form of the
coupling functions, no change is expected apart from the form of
$\delta_\xi$, $\epsilon_s$ and indices $\epsilon_5$ and
$\epsilon_6$,
\begin{equation}
\centering
\label{dxiB2}
\delta_\xi\simeq\frac{m(1-\beta)}{3(m-1)}\frac{\xi(\phi)}{h(\phi)}\kappa^4V(\phi),\,
\end{equation}
\begin{equation}
\centering
\label{esB2}
\epsilon_s\simeq\left(\frac{(1-\beta)\gamma\kappa\phi}{m-1}\right)^2+\frac{1-\beta}{(m-1)h(\phi)}\left(\frac{\Lambda\gamma\kappa\phi\cos(\gamma\kappa\phi+\theta)}{2}-\frac{4m\xi(\phi)\kappa^4V(\phi)}{3h(\phi)}\right),\,
\end{equation}
\begin{equation}
\centering
\epsilon_5=\frac{(1-\beta) \left(8\lambda  m \kappa^4V(\phi ) (\kappa  \phi )^m \csc ^2(\gamma  \kappa  \phi +\theta )-3 \gamma  \Lambda ^2 \kappa\phi  \cot (\gamma  \kappa  \phi +\theta )\right)}{ 16 (1-\beta ) \lambda  m \kappa^4V(\varphi ) (\kappa  \varphi )^m \csc ^2(\gamma  \kappa  \phi +\theta )-6 \Lambda ^2 (m-1)},\,
\end{equation}
\begin{equation}
\centering
\epsilon_6=\frac{(1-\beta ) \left(3 \gamma  \Lambda ^2 (m-1) \kappa\phi  \sin (2 (\gamma  \kappa  \phi +\theta ))-16 (1-\beta ) \lambda  m (\kappa  \phi )^m \left(\kappa^4V(\phi ) (m-\gamma  \kappa  \phi  \cot (\gamma  \kappa  \phi +\theta ))+\kappa\varphi \kappa^3 V'(\phi )\right)\right)}{4 (m-1) \left(3 \Lambda ^2 (m-1) \sin ^2(\gamma  \kappa  \phi +\theta )-8 (1-\beta )  \lambda  m \kappa^4V(\phi ) (\kappa  \phi )^m\right)}.\,
\end{equation}
The same applies to the expressions of the values for the scalar
field. Thus, designating  the exact same values of parameters as
before, with the only exception being parameters $\Lambda$ and
$V_0$ which are now assigned the values $\Lambda=0.4$ and
$V_0=10^{-6}$ leads to compatible results as $n_s=0.965226$,
$n_T=-0.00001854$ and $r=0.0001483$ are acceptable. Moreover, the
non-Gaussianities of the model are expected to obtain the values
$f_{NL}^{eq}=0.0818867$ and
$\left<\mathcal{R}^3\right>=0.0769239$, with a corresponding
skewness $\mathcal{S}\sim0.00001654$. This is a completely
different expected amount of deviation from previously but still
very small as expected.

Unfortunately, even this case is not viable since a single
approximation is violated. We mention that
$h''\dot\phi^2\sim\mathcal{O}(10^{-18})$ while $H\dot
h\sim\mathcal{O}(10^{-12})$, hence the form of Hubble's derivative
in Eq. (\ref{motion2B}) is not justifiable.

\subsection{Model III: Power-Law and Exponential Couplings}

\begin{figure}[h!]
\centering
\includegraphics[width=20pc]{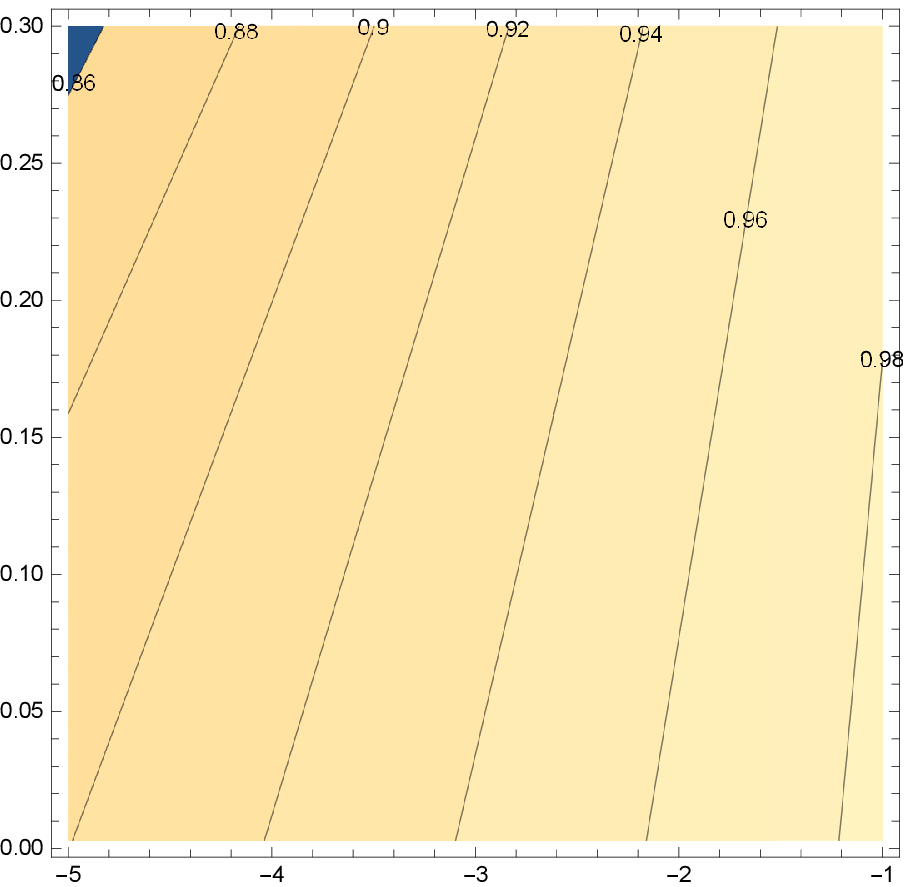}
\includegraphics[width=20pc]{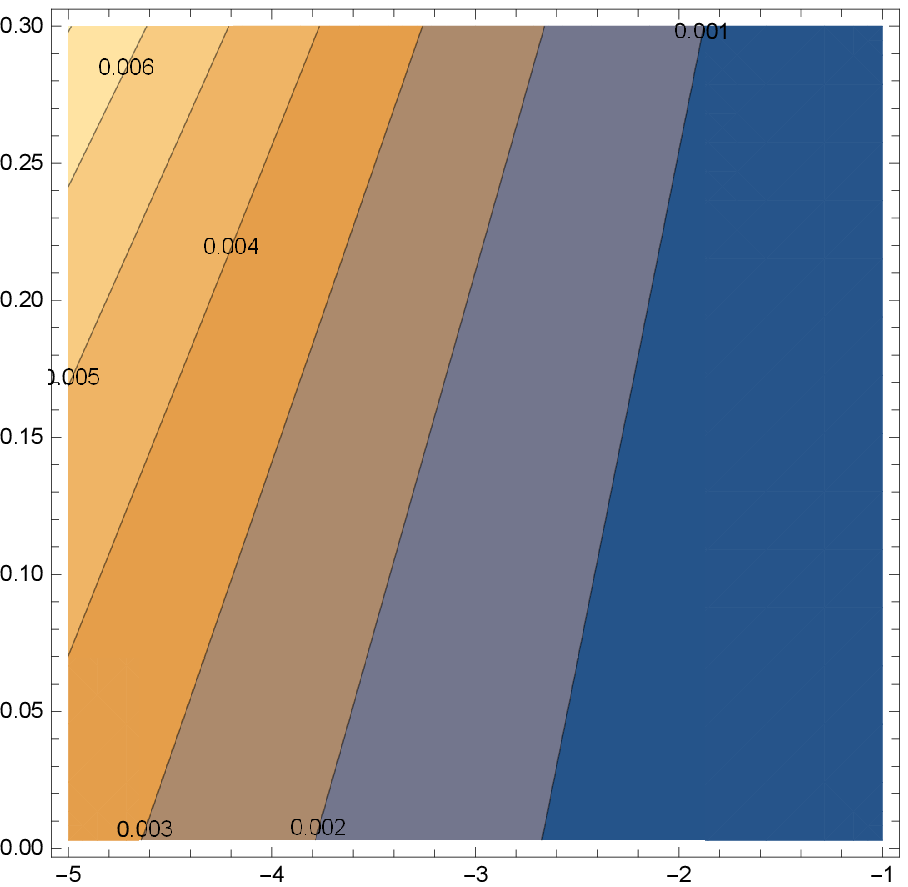}
\caption{Contour plots  of the spectral index of scalar
perturbations $n_S$ (left) and the tensor-to-scalar ratio $r$
(right) depending on parameters $\beta$ and $\gamma$. Their values
range from [0.003, 0.3] and [-5,-1] respectively. It is clear that
the both objects depend strongly on each parameter.} \label{plot2}
\end{figure}

In this case, we shall examine a model reminiscing the first one
presented, but this time the Ricci coupling will be generalized.
Let,
\begin{equation}
\centering
\label{h3}
h(\phi)=\Lambda (\kappa\phi)^m,\,
\end{equation}
\begin{equation}
\centering
\label{xi3}
\xi(\phi)=\lambda e^{\gamma\kappa\phi}.\,
\end{equation}
This choice is made similarly due to the simple ratios in the
coupling functions. Furthermore, we shall use the following forms
of the equations of motion,
\begin{equation}
\centering
\label{motion1C}
H^2\simeq\frac{\kappa^2V}{3h\left(1+(1-\beta)\frac{h'}{h}\frac{\xi'}{\xi''}\right)},\,
\end{equation}
\begin{equation}
\centering
\label{motion2C}
\dot H\simeq\frac{H^2}{2}(1-\beta)^2\frac{\xi'}{\xi''}\left(\frac{h'}{h}-\frac{h''}{h}\frac{\xi'}{\xi''}\right),\,
\end{equation}
\begin{equation}
\centering
\label{motion3C}
V'-6H^2\frac{h'}{\kappa^2}\simeq0.\,
\end{equation}
Here, it is convenient to make use of the positive sign of
slow-roll index $\epsilon_1$. Also, we shall use the more
simplified differential equation for the scalar potential since,
as mentioned before, leads to functionally more elegant results.
In fact, the scalar potential now is,
\begin{equation}
\centering
\label{potC}
V(\phi)=V_0(m(1-\beta)+\gamma\kappa\phi)^{2m},\,
\end{equation}
where similarly, the integration constant $V_0$ has again mass
dimensions [m]$^4$. In addition, the auxiliary parameters and
slow-roll indices are written as,
\begin{equation}
\centering
\label{dXC}
\delta_X=\frac{\omega(1-\beta)^2}{2\gamma^2h(\phi)},\,
\end{equation}
\begin{equation}
\centering
\label{dxiC}
\delta_\xi\simeq\frac{1-\beta}{3}\frac{\xi(\phi)}{h^2(\phi)}\frac{\kappa^2V(\phi)}{\left(1+\frac{m(1-\beta)}{\gamma\kappa\phi}\right)},\,
\end{equation}
\begin{equation}
\centering
\label{esC}
\epsilon_s\simeq-\frac{m(1-\beta)^2}{2\gamma\kappa\phi}\left(1-\frac{m-1}{\gamma\kappa\phi}\right)+(1-\beta)\left(\frac{m}{\gamma\kappa\phi}-\frac{4\xi(\phi)\kappa^4V(\phi)}{3h^2(\phi)\left(1+\frac{m(1-\beta)}{\gamma\kappa\phi}\right)}\right),\,
\end{equation}
\begin{equation}
\centering
\label{index1C}
\epsilon_1\simeq-\frac{m(1-m-\gamma\kappa\phi)}{2}\left(\frac{1-\beta}{\gamma\kappa\phi}\right)^2,\,
\end{equation}
\begin{equation}
\centering
\label{index2C}
\epsilon_2=\beta,\,
\end{equation}
\begin{equation}
\centering
\label{index3C}
\epsilon_3=\frac{m(1-\beta)}{2\gamma\kappa\phi},\,
\end{equation}
\begin{widetext}
\[
\centering
\label{index5C}
\epsilon_5=\frac{(\beta -1) \left(3 \Lambda ^2 m (\kappa  \phi )^{2 m} ((\beta -1) m-\gamma  \kappa  \phi )+2\lambda(2\gamma \kappa\phi)^2  \kappa ^4 V(\phi ) e^{\gamma  \kappa  \phi }\right)}{2 \gamma  \kappa  \phi  \left(8 (\beta -1) \gamma  \kappa\phi \lambda  \kappa^4V(\phi ) e^{\gamma  \kappa  \phi }-3 \Lambda ^2 (\kappa  \phi )^{2 m} ((\beta -1) m-\gamma  \kappa  \phi )\right)},\,
\]
\end{widetext}
\begin{equation}
\centering
\label{index6C}
\epsilon_6\simeq\epsilon_5.\,
\end{equation}
This choice leads to a second order polynomial slow-roll index
$\epsilon_1$ hence two values of the scalar field at the end of
inflation appear, once the index $\epsilon_1$ becomes of order
$\mathcal{O}(1)$. These values are showcased below,
\begin{equation}
\centering
\label{scalarf_C}
\phi_f=\frac{(1-\beta) \sqrt{m} \left( (1-\beta) \sqrt{m} \pm\sqrt{\left(\beta ^2-2 \beta -7\right) m+8}\right)}{4 \gamma  \kappa }.\,
\end{equation}
In order to derive the expression of the initial value of the
scalar field from Eq. (\ref{efolds}). we shall assume the first
case with the plus sign. As a result, the formula for the initial
value is,
\begin{equation}
\centering
\label{scalariC}
\phi_i=\frac{(1-\beta ) \left( (1-\beta ) m-4 N -\sqrt{m} \sqrt{\left(\beta ^2-2 \beta -7\right) m+8}\right)}{4 \gamma  \kappa }.\,
\end{equation}
Assigning the following values ($\omega$, $\Lambda$, $\lambda$,
$V_0$, $N$, $m$, $\gamma$, $\beta$)=(1, 1650, $10^{-4}$,
$10^{-6}$, 60, 0.5, -2, 0.0001) produces viable results as
$n_S=0.963545$, $n_T=0.00006998$ and $r=0.00055759$ are compatible
with the observations. Furthermore, for the non-Gaussianities,
$f_{NL}^{eq}$=0.0136957 and
$\left<\mathcal{R}^3\right>=1.25\cdot10^{-17}$, extremely small
values with a corresponding predicted deviation of
$\mathcal{S}\sim4.784\cdot10^{-25}$, which in turn implies that
there exists no effective deviation. Concerning the slow-roll
indices, we mention that the slow-roll criteria are indeed true as
$\epsilon_1=-0.004114$, $\epsilon_3=-0.004149$,
$\epsilon_4=0.005674$, $\epsilon_5=-0.004149=\epsilon_6$ and also,
the auxiliary parameters for evaluating the non-Gaussianities
obtain the values $\delta_\xi=-1.679\cdot10^{-43}$,
$\delta_X=-3.0603\cdot10^{-14}$, $\eta=0.0329975$,
$\epsilon_s=-0.00003484$ and $s=-1.1716\cdot10^{-33}$ which are
extremely small values. Finally, for the scalar field we note that
$\phi_i=30.122$ while $\phi_f=0.125$ which shows a decrease with
time. The inflationary phenomenology of the model at hand is
presented in Figs. \ref{plot2} and \ref{parplot2} for some ranges
of values of the free parameters.
\begin{figure}[h!]
\centering
\includegraphics[width=20pc]{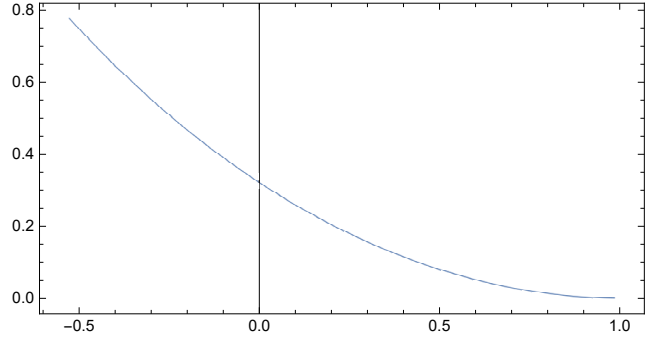}
\caption{Parametric plot of spectral index of scalar perturbations
$n_S$ (x axis) and the tensor-to-scalar ratio $r$ (y axis)
depending on parameters $\beta$ and $\gamma$. Their values range
from [0.0003, 0.001] and [-75,-1] respectively. There exists a
one-on one correlation between these quantities and a variety of
possible set of values. The compatible with the observations
results are on the bottom right.} \label{plot3}
\end{figure}
This model has interesting dynamics, since altering some values
yields different characteristics. For instance, we mention that
$\gamma$ must be negative, otherwise the value of the scalar field
during the first horizon crossing becomes negative and
subsequently, the spectral indices obtain an imaginary part where
in certain cases is even grater, in absolute value, than the real
part. The same applies to the exponent $m$ for each value,
positive or negative, small or large. Only when $m=0.5$ the model
yields real numbers and in fact, changing both $m$ and $\gamma$
does not negate the appearance of complex numbers. Moreover, the
constant-roll parameter $\beta$ influences greatly the spectral
index of scalar perturbations and in second terms the
tensor-to-scalar ratio as designating the value $\beta=0.001$
leads to $n_S=0.961747$ and $r=0.000617343$, so $\beta$ plays
significant role. This particular parameter can also obtain
negative values and still lead to viable results. In contrast,
$\Lambda$ alters only the spectral index of primordial curvature
perturbations as taking the value $\Lambda=1450$ leads to
$n_s=0.964103$ and $\Lambda=450$ to $n_S=0.9706$. Changing the
sign of $\Lambda$ also influences the spectral index in the same
manner. Lastly, decreasing $\gamma$ leads to a subsequent decrease
in the spectral index of scalar perturbations while simultaneously
keeping the tensor-to-scalar invariant. The tensor-to-scalar ratio
is very strictly assigned as only the constant-roll parameter
changes is and not significantly as we showed previously. The
decisive factor is the exponent $m$ and the fact that only one
value leads to not only real numbers, but also a viable
tensor-to-scalar ratio is fascinating.

Concerning the non-Gaussianities, $\beta$ alters only the
numerical value and not the order of magnitude of both the non
linear term and the three-point correlator. On the other hand,
decreasing $\gamma$, for instance to the value $\gamma=20$
increases the order of magnitude of the three-point correlator
approximately three orders, while leaving invariant non-linear
term. Finally, as expected, parameters $\Lambda$ and $V_0$
influence the order of magnitude of the three point correlator.
Again, this is attributed to Hubble's parameter and the connection
it has with the three-point correlator.

This description seems to be correct since all the approximations
made actually do apply, but in certain cases marginally, We
mention that for the slow-roll conditions, $\dot
H/H^2\sim\mathcal{O}(10^{-3})$ and additionally,
$X/V\sim\mathcal{O}(10^{-6})$ so indeed they apply. Moreover, for
the equations of motion (\ref{motion1C}) and (\ref{motion3C}) we
note that $24\dot\xi H^3\sim\mathcal{O}(10^{-47})$ and
$24\xi'H^4\sim\mathcal{O}(10^{-46})$ so it justifies why the
string corrections where omitted in equations (\ref{motion1}) and
(\ref{motion3}) respectively. Also, $h'\sim\mathcal{O}(10^2)$
while $\xi'/\xi''\sim\mathcal{O}(10^{-1})$ so the extra
approximation does indeed apply. Lastly, for Eq. (\ref{motion2C}),
we note that $16\dot\xi H\dot H\sim\mathcal{O}(10^{-50})$ while
the terms used in our approach, meaning $H\dot h$ and
$h''\dot\phi^2$ are of order $\mathcal{O}(10^{-7})$ and
$\mathcal{O}(10^{-9})$ respectively so it justifies why the string
corrective term was neglected. However, the kinetic term
$\dot\phi^2$ which was discarded as well happens to be of order
$\mathcal{O}(10^{-10})$ so it is quite close to the term
$h''\dot\phi^2$. However, we can thing such term as a first
correction of the same model but with an expression for Hubble's
derivative exactly as in the previous model II. Therefore, adding
the term $h''\dot\phi^2$ in order to get a better understanding
and treating it as a first order corrective term and not as
dominant as the term $H\dot h$ makes sense. For the sake of
completeness we mention that one can overcome such obstacle when
working in the non-canonical case. Assuming for instance that
$\omega=0.001$ and changing the constant-roll parameter to the
value $\beta=-0.005$ leads again to viable results while keeping
the orders of magnitude of each element exactly as before, with
the only exception being the kinetic term. Then these terms become
truly dominant compared to the rest.

Let us now compare the previous model with the slow-roll case. As
was the case with model I, letting $\beta=0$ and changing
$\epsilon_2$ leads to, as expected, the exact same potential.
Changing only one parameter, $\gamma=-20$, leads to compatible
results as $n_S=0.966737$, $n_T=0.0000703$, $r=0.0005602$,
$f_{NL}^{eq}=0.0138935$,
$\left<\mathcal{R}^3\right>=1.26423\cdot10^{-15}$ and
$\mathcal{S}\sim1.51686\cdot10^{-22}$ Hence, both approaches yield
viable phenomenology. This is expected since $\beta$ obtains small
values, hence $\ddot\phi$ can in fact be discarded in Eq.
(\ref{motion2}) as was the case with the slow-roll condition.
However it is worth inspecting the constant-roll case, even if it
produces similar results since both conditions are intrinsically
different.

\section{Panorama of All Possible Approximations During the Inflationary Era and Their Validity}

In this section, we summarize all the possible approaches one can
use when studying the inflationary era of a non-minimally coupled
model supplemented with string correction terms under the
constant-roll condition. Each one is accompanied by the
approximations that must apply during the first horizon crossing.
As it was shown previously, there exist multiple different
approaches in each separate equation of motion one can work with,
each one promising and capable of leading to functional
expressions. Despite the choice, one must always check whether the
approximations made in the model apply, simply because even a
single invalid approximation renders the model false, even if it
is capable of producing viable results.

Before elaborating any further, it is worth mentioning that no
matter the choices for the equations, there exist two
approximations which must apply no matter the case. Previously,
during our analysis in the viability of certain models, we were
free to change our approach if stumbled on a violating condition
although this is indicative of the flexibility this formalism
contains. However, no matter the choice, the slow-roll conditions
must always apply, meaning that in every model, the following
relations must be respected,
\begin{align}
\centering
\dot H& \ll H^2,& \frac{1}{2}\omega\dot\phi^2&\ll V.\,
\end{align}
In other words, these are the main approximations that must apply
no matter the case while the rest approximations are auxiliary, in
order to make the system of equations solvable. Consequently, it
is worth inspecting them separately in each equation of motion.

For the differential equation of the scalar potential, we mention
that there exist two possible ways one can work with easily, as
shown below,
\begin{equation}
\centering
\label{Vdif1}
V'+3H^2\left(\omega(1-\beta)(1+\frac{\beta}{3})\frac{\xi'}{\xi''}-2\frac{h'}{\kappa^2}\right)\simeq0,\,
\end{equation}
\begin{equation}
\centering
\label{Vdif2}
V'-6H^2\frac{h'}{\kappa^2}\simeq0.
\end{equation}
Working with the full expression as Eq. (\ref{motion6a}) is always
a possibility but adding more terms in the differential equation
which are also less dominant make the form of the scalar potential
perplexed for no apparent reason, so it is convenient to discard
them completely. For the corresponding approximations, one must
check whether the following correlations hold,
\begin{align}
\centering
24\xi'H^4&\ll V',&24\kappa\xi'H^2&\ll \kappa\xi'/\xi''-2h',&\kappa\xi'/\xi''&\ll h'.\,
\end{align}
When a correlation contains multiple expressions on each side, one
must always examine each term from the right separately which each
term on the lest. Moreover, what matters is obviously the order of
magnitude and not the sign of each term. Lastly, the last
condition may be violated when one works with Eq. (\ref{Vdif1}).

For the expression of Hubble's parameter, one has the ability to
choose from three possible expressions,
\begin{equation}
\centering
\label{H1}
H^2\simeq\frac{\kappa^2V}{3h\left(1+(1-\beta)\frac{h'}{h}\frac{\xi'}{\xi''}\right)},\,
\end{equation}
\begin{equation}
\centering
\label{H2}
H^2\simeq\frac{\kappa^2V}{3h},\,
\end{equation}
\begin{equation}
\centering
\label{H3}
H^2\simeq\frac{\kappa^2V}{3(1-\beta)h'}\frac{\xi''}{\xi'}.\,
\end{equation}
As it was demonstrated in the previous model, there are cases in
which one form is more preferable that the rest, so one must
choose wisely. Also, the corresponding assumptions that must apply
are,
\begin{align}
\centering
14\kappa^2\dot\xi H^3&\ll \kappa^2V-3H^2h-3H\dot h,& H\dot h&\ll hH^2,&hH^2&\ll H\dot h.\,
\end{align}
The last conditions may seem contradictory but we note that the
second refers to equation (\ref{H2}) while the third to (\ref{H3})
so they make sense. In addition, the first approximation must
apply always, in each form of Hubble's parameter since in every
single approach the string corrections were discarded.

Lastly, the expression for  Hubble's derivative is the most
crucial as different approaches lead to different characteristics
and therefore different values for the scalar field during the
first horizon crossing, leading to subsequent different values for
the observed quantities and the predicted non-Gaussianities. Thus,
different choices lead to different phenomenology. When working
under the constant-roll condition, one sees that the term
$\ddot\phi h'$ can also be included in the equation, a feature
which can not be applies in the slow-roll case. Thus, there exist
six inherently different forms, as presented below,
\begin{equation}
\centering
\label{Hdot1}
\dot H\simeq\frac{H^2}{2}(1-\beta)^2\frac{h'}{h}\frac{\xi'}{\xi''},\,
\end{equation}
\begin{equation}
\centering
\label{Hdot2}
\dot H\simeq-\frac{H^2}{2}(1-\beta)^2\frac{h''}{h}\left(\frac{\xi'}{\xi''}\right)^2,\,
\end{equation}
\begin{equation}
\centering
\label{Hdot3}
\dot H\simeq-\frac{H^2}{2h}\kappa^2\omega(1-\beta)^2\left(\frac{\xi'}{\xi''}\right)^2,\,
\end{equation}
\begin{equation}
\centering
\label{Hdot4}
\dot H\simeq\frac{H^2}{2}(1-\beta)^2\frac{\xi'}{\xi''}\left(\frac{h'}{h}-\frac{h''}{h}\frac{\xi'}{\xi''}\right),\,
\end{equation}
\begin{equation}
\centering
\label{Hdot5}
\dot H\simeq\frac{H^2}{2}(1-\beta)^2\frac{\xi'}{\xi''}\left(\frac{h'}{h}-\frac{\kappa^2\omega}{h}\frac{\xi'}{\xi''}\right),\,
\end{equation}
\begin{equation}
\centering
\label{Hdot6}
\dot H\simeq-\frac{H^2}{2}(1-\beta)^2\left(\frac{\xi'}{\xi''}\right)^2\left(\frac{\kappa^2\omega}{h}+\frac{h''}{h}\right).\,
\end{equation}
These are all the possible forms of Hubble's derivative. There
exists also the possibility of keeping every term of
(\ref{motion2}) apart from the string correction and for with such
model. This is attributed to the constant-roll condition which now
enables us to work with a functional form of the term $\ddot\phi$.
However this case is very intricate and if it is solvable, it
might by only for the simplest case like a power-law expression.
Such case could correspond to the first model as well where the
linear choice for Ricci's coupling function discarded naturally
the second derivative. Nevertheless, it is wise to present the
simplified choices as well since in many models as showcased
previously, certain terms are not so dominant and in fact can be
neglected. Let us now proceed with the corresponding
approximations for each model separately.

Working with Eq. (\ref{Hdot1}), the results must respect the
following conditions,
\begin{align}
\centering
16\kappa^2\dot\xi H &\ll 2h,& (\kappa^2\omega+h'')\dot\phi&\ll (1-\beta)Hh'.\,
\end{align}
In a similar way, choosing Eq. (\ref{Hdot2}), we get the following
conditions,
\begin{align}
\centering
16\kappa^2\dot\xi H&\ll 2h,&\kappa^2\omega&\ll h'',&(1-\beta)Hh'&\ll \kappa^2h''\dot\phi.\,
\end{align}
In the case of  Eq. (\ref{Hdot3}), we have,
\begin{align}
\centering
16\kappa^2\dot\xi H&\ll 2h,&h''&\ll\kappa^2\omega,&(1-\beta)Hh'&\ll \kappa^2\dot\phi.\,
\end{align}
If one were to work with Eq. (\ref{Hdot4}), then,
\begin{align}
\centering
16\kappa^2\dot\xi H&\ll 2h,&\kappa^2\omega\dot\phi&\ll(1-\beta)Hh'-h''\dot\phi.
\end{align}
The choice of Eq. (\ref{Hdot5}) demands the following conditions
hold true,
\begin{align}
\centering
16\kappa^2\dot\xi H&\ll 2h,&h''\dot\phi&\ll(1-\beta)Hh'-\kappa^2\omega\dot\phi .\,
\end{align}
Finally, the last choice of Eq. (\ref{Hdot6}) leads to the
following expressions,
\begin{align}
\centering
16\kappa^2\dot\xi H&\ll 2h,&(1-\beta)Hh'&\ll(\kappa^2\omega+h'')\dot\phi.\,
\end{align}
The first condition is the same in each choice and it simply
states that we chose Hubble's derivative from the term $2h\dot H$
and not the string corrections. This was also done in the case of
Hubble's parameter itself were the string correction, which was
proportional toy $H^3$ was discarded in each case. These are all
the possible approaches one must follow. Each condition written
here for each case must be respected and once again, we note that
the order of magnitude matters and not the sign itself. If these
assumptions hold during the first horizon crossing, then the model
is indeed viable.

\section{Conclusions}

In this paper, we studied the viability of a non-minimally coupled
Einstein-Gauss-Bonnet gravity with the constant-roll condition
holding true during the inflationary era. We presented a new
framework, in which constraints of the velocity of the primordial
tensor perturbations of the metric, which are essential in order
for a model not to be at variance with the physical world. This
constraint leads to a decrease in the degrees of freedom and
subsequently to interesting dynamics. Here, we assumed for
convenience that the scalar potential, which is in fact present in
this framework, is not freely designated but on the contrary is
derivable from the equations of motion once the scalar functions
coupled to the Ricci scalar and Gauss-Bonnet invariant are
specified. However, a completely different approach where one of
the coupling functions is derived from the equations of motion is
also feasible but was not studied further. We presented all the
possible approximations which can be implemented and showed that
each one is capable of producing different phenomenology, in
certain cases even elegantly. The main result that is extracted
from this paper is that even though the constant-roll is
intrinsically different from the slow-roll assumption, both
approaches are more than capable of producing viable phenomenology
during the inflationary era, however, the non-Gaussianities
predicted are small, a feature that we did not anticipate, since
the constant-roll evolution is known to produce larger
non-Gaussianities compared to the slow-roll case.

\section*{Appendix}

We introduce a simple and elegant way of deriving a functional
expression of $\xi(\phi)$. In each approach, it was shown that
this particular functions appears in ratios, hence simplifying the
ratios facilitates our study greatly. As a result, the first
slow-roll index, the $e$-foldings number and the initial and final
value of the scalar field have simple forms. The only requirement
is that the coupling scalar function is at least three times
differentiable. This particular formalism is shown below,
\begin{equation}
\centering
\label{eqxi'}
\xi'(\phi)=\kappa\lambda e^{\int{\kappa X[\phi]d\phi}},\,
\end{equation}
where $\lambda$ is a dimensionless parameter and $X[\phi]$ is
dimensionless arbitrary expression depending on the scalar field.
This form was chosen simply because by differentiation with
respect to the scalar field, we end up with the following
expression,
\begin{equation}
\centering
\xi''=\kappa X[\phi]\xi'.\,
\end{equation}
Thus, the ratio $\xi'/\xi''$ which appears in all equations is
replaced by the term $X[\phi]$. Choosing appropriately this term
leads to an easy phenomenology. One can choose to work with such
term in order to find an appealing and functional formula for the
initial value of the scalar field $\phi$ and then later derive the
expression of the Gauss-Bonnet coupling scalar function by simply
integrating Eq. (\ref{eqxi'}).

The same applies to the Ricci coupling function as well. Recall
that in certain approaches, the function participates in the form
of a ratio $h'/h$, so extending the previous formalism, one can
work with the form,
\begin{equation}
\centering
\label{eqh}
h(\phi)=\Lambda e^{\int{\kappa Y[\phi]d\phi}},\,
\end{equation}
where $\Lambda$ a dimensionless constant and $ Y[\phi]$ a
dimensionless arbitrary function of the scalar field. As was with
the previous case,
\begin{equation}
\centering
h'=\kappa Y[\phi]h.\,
\end{equation}
Thus, specifying $Y[\phi]$ determines the ratio for Ricci's
coupling function while simultaneously designating its form. Note
that changing $\Lambda$ or $\lambda$ does not affect the form of
the ratio of the coupling functions, or in this case $X[\phi]$ and
$Y[\phi]$. Lastly, concerning a different ratio of the Ricci
coupling function, the same approach still applies, although in
this case the resulting equation is different. Now,
\begin{equation}
\centering
\frac{h''}{h}=(\kappa Y[\phi])^2+\kappa Y'[\phi],\,
\end{equation}
so choosing an appropriate function which satisfies the above
equation can simplify the results greatly. For instance, a linear
choice for $Y[\phi]$ results in an exponential of a trigonometric
function which makes this ratio constant, as was the case with the
second model.

In view of this formalism, one is capable of working the other way
around. Instead of guessing the coupling functions, the user
chooses what they think of as an appropriate feature for the ratio
and once this designating yields functional or elegant results,
work backwards in order to find the coupling scalar functions
responsible for such results. In the non minimal case, the task is
more difficult since these two forms, $X[\phi]$ and $Y[\phi]$
could have constructive or destructive behavior so choosing wisely
both is the key to elegant phenomenology. This is because choosing
appropriately both forms could simplify greatly the results. The
choice is obviously up to the reader.


\begin{thebibliography}{99}



\bibitem{Hwang:2005hb}
  J.~c.~Hwang and H.~Noh,
  Phys.\ Rev.\ D {\bf 71} (2005) 063536
  doi:10.1103/PhysRevD.71.063536
  [gr-qc/0412126].


\bibitem{Nojiri:2006je}
  S.~Nojiri, S.~D.~Odintsov and M.~Sami,
  Phys.\ Rev.\ D {\bf 74} (2006) 046004
  doi:10.1103/PhysRevD.74.046004
  [hep-th/0605039].




\bibitem{Cognola:2006sp}
  G.~Cognola, E.~Elizalde, S.~Nojiri, S.~Odintsov and S.~Zerbini,
  Phys.\ Rev.\ D {\bf 75} (2007) 086002
  doi:10.1103/PhysRevD.75.086002
  [hep-th/0611198].



\bibitem{Nojiri:2005vv}
  S.~Nojiri, S.~D.~Odintsov and M.~Sasaki,
  Phys.\ Rev.\ D {\bf 71} (2005) 123509
  doi:10.1103/PhysRevD.71.123509
  [hep-th/0504052].


\bibitem{Nojiri:2005jg}
  S.~Nojiri and S.~D.~Odintsov,
  Phys.\ Lett.\ B {\bf 631} (2005) 1
  doi:10.1016/j.physletb.2005.10.010
  [hep-th/0508049].







\bibitem{Satoh:2007gn}
  M.~Satoh, S.~Kanno and J.~Soda,
  Phys.\ Rev.\ D {\bf 77} (2008) 023526
  doi:10.1103/PhysRevD.77.023526
  [arXiv:0706.3585 [astro-ph]].



\bibitem{Bamba:2014zoa}
  K.~Bamba, A.~N.~Makarenko, A.~N.~Myagky and S.~D.~Odintsov,
  JCAP {\bf 1504} (2015) 001
  doi:10.1088/1475-7516/2015/04/001
  [arXiv:1411.3852 [hep-th]].


\bibitem{Yi:2018gse}
  Z.~Yi, Y.~Gong and M.~Sabir,
  Phys.\ Rev.\ D {\bf 98} (2018) no.8,  083521
  doi:10.1103/PhysRevD.98.083521
  [arXiv:1804.09116 [gr-qc]].


\bibitem{Guo:2009uk}
  Z.~K.~Guo and D.~J.~Schwarz,
  Phys.\ Rev.\ D {\bf 80} (2009) 063523
  doi:10.1103/PhysRevD.80.063523
  [arXiv:0907.0427 [hep-th]].


\bibitem{Guo:2010jr}
  Z.~K.~Guo and D.~J.~Schwarz,
  Phys.\ Rev.\ D {\bf 81} (2010) 123520
  doi:10.1103/PhysRevD.81.123520
  [arXiv:1001.1897 [hep-th]].


\bibitem{Jiang:2013gza}
  P.~X.~Jiang, J.~W.~Hu and Z.~K.~Guo,
  Phys.\ Rev.\ D {\bf 88} (2013) 123508
  doi:10.1103/PhysRevD.88.123508
  [arXiv:1310.5579 [hep-th]].



\bibitem{Kanti:2015pda}
  P.~Kanti, R.~Gannouji and N.~Dadhich,
  Phys.\ Rev.\ D {\bf 92} (2015) no.4,  041302
  doi:10.1103/PhysRevD.92.041302
  [arXiv:1503.01579 [hep-th]].


\bibitem{vandeBruck:2017voa}
  C.~van de Bruck, K.~Dimopoulos, C.~Longden and C.~Owen,
  arXiv:1707.06839 [astro-ph.CO].



\bibitem{Kanti:1998jd}
  P.~Kanti, J.~Rizos and K.~Tamvakis,
  Phys.\ Rev.\ D {\bf 59} (1999) 083512
  doi:10.1103/PhysRevD.59.083512
  [gr-qc/9806085].




\bibitem{Pozdeeva:2020apf}
  E.~O.~Pozdeeva, M.~R.~Gangopadhyay, M.~Sami, A.~V.~Toporensky and S.~Y.~Vernov,
  arXiv:2006.08027 [gr-qc].

\bibitem{Fomin:2020hfh}
  I.~Fomin,
  arXiv:2004.08065 [gr-qc].

\bibitem{DeLaurentis:2015fea}
  M.~De Laurentis, M.~Paolella and S.~Capozziello,
  Phys.\ Rev.\ D {\bf 91} (2015) no.8,  083531
  doi:10.1103/PhysRevD.91.083531
  [arXiv:1503.04659 [gr-qc]].


\bibitem{Chervon:2019sey}
  S.~Chervon, I.~Fomin, V.~Yurov and A.~Yurov,
  doi:10.1142/11405



\bibitem{Nozari:2017rta}
  K.~Nozari and N.~Rashidi,
  Phys.\ Rev.\ D {\bf 95} (2017) no.12,  123518
  doi:10.1103/PhysRevD.95.123518
  [arXiv:1705.02617 [astro-ph.CO]].




\bibitem{Odintsov:2018zhw}
  S.~D.~Odintsov and V.~K.~Oikonomou,
  Phys.\ Rev.\ D {\bf 98} (2018) no.4,  044039
  doi:10.1103/PhysRevD.98.044039
  [arXiv:1808.05045 [gr-qc]].


  \bibitem{Kawai:1998ab}
  S.~Kawai, M.~a.~Sakagami and J.~Soda,
  Phys.\ Lett.\ B {\bf 437}, 284 (1998)
  doi:10.1016/S0370-2693(98)00925-3
  [gr-qc/9802033].


\bibitem{Yi:2018dhl}
  Z.~Yi and Y.~Gong,
  Universe {\bf 5} (2019) no.9,  200
  doi:10.3390/universe5090200
  [arXiv:1811.01625 [gr-qc]].


\bibitem{vandeBruck:2016xvt}
  C.~van de Bruck, K.~Dimopoulos and C.~Longden,
  Phys.\ Rev.\ D {\bf 94} (2016) no.2,  023506
  doi:10.1103/PhysRevD.94.023506
  [arXiv:1605.06350 [astro-ph.CO]].


\bibitem{Kleihaus:2019rbg}
  B.~Kleihaus, J.~Kunz and P.~Kanti,
  arXiv:1910.02121 [gr-qc].





\bibitem{Bakopoulos:2019tvc}
  A.~Bakopoulos, P.~Kanti and N.~Pappas,
  Phys.\ Rev.\ D {\bf 101} (2020) no.4,  044026
  doi:10.1103/PhysRevD.101.044026
  [arXiv:1910.14637 [hep-th]].


\bibitem{Maeda:2011zn}
  K.~i.~Maeda, N.~Ohta and R.~Wakebe,
  Eur.\ Phys.\ J.\ C {\bf 72} (2012) 1949
  doi:10.1140/epjc/s10052-012-1949-6
  [arXiv:1111.3251 [hep-th]].






\bibitem{Bakopoulos:2020dfg}
  A.~Bakopoulos, P.~Kanti and N.~Pappas,
  arXiv:2003.02473 [hep-th].


\bibitem{Ai:2020peo}
W.~Ai,
[arXiv:2004.02858 [gr-qc]].



\bibitem{Odintsov:2019clh}
  S.~D.~Odintsov and V.~K.~Oikonomou,
  Phys.\ Lett.\ B {\bf 797} (2019) 134874
  doi:10.1016/j.physletb.2019.134874
  [arXiv:1908.07555 [gr-qc]].



\bibitem{Oikonomou:2020oil}
V.~K.~Oikonomou and F.~P.~Fronimos,
[arXiv:2007.11915 [gr-qc]].

\bibitem{Odintsov:2020xji}
S.~D.~Odintsov, V.~K.~Oikonomou and F.~P.~Fronimos,
Annals Phys. \textbf{420} (2020), 168250
doi:10.1016/j.aop.2020.168250 [arXiv:2007.02309 [gr-qc]].



\bibitem{Oikonomou:2020sij}
V.~K.~Oikonomou and F.~P.~Fronimos,
[arXiv:2006.05512 [gr-qc]].



\bibitem{Odintsov:2020zkl}
S.~D.~Odintsov and V.~K.~Oikonomou,
Phys. Lett. B \textbf{805} (2020), 135437
doi:10.1016/j.physletb.2020.135437 [arXiv:2004.00479 [gr-qc]].


\bibitem{Odintsov:2020sqy}
S.~D.~Odintsov, V.~K.~Oikonomou and F.~P.~Fronimos,
[arXiv:2003.13724 [gr-qc]].




\bibitem{Odintsov:2020mkz}
S.~D.~Odintsov, V.~K.~Oikonomou, F.~P.~Fronimos and
S.~A.~Venikoudis,
Phys. Dark Univ. \textbf{30} (2020), 100718
doi:10.1016/j.dark.2020.100718 [arXiv:2009.06113 [gr-qc]].


\bibitem{Easther:1996yd}
  R.~Easther and K.~i.~Maeda,
  Phys.\ Rev.\ D {\bf 54} (1996) 7252
  doi:10.1103/PhysRevD.54.7252
  [hep-th/9605173].

\bibitem{Antoniadis:1993jc}
  I.~Antoniadis, J.~Rizos and K.~Tamvakis,
  Nucl.\ Phys.\ B {\bf 415} (1994) 497
  doi:10.1016/0550-3213(94)90120-1
  [hep-th/9305025].

\bibitem{Antoniadis:1990uu}
I.~Antoniadis, C.~Bachas, J.~R.~Ellis and D.~V.~Nanopoulos,
Phys.\ Lett.\ B \textbf{257} (1991), 278-284
doi:10.1016/0370-2693(91)91893-Z




\bibitem{Kanti:1995vq}
P.~Kanti, N.~Mavromatos, J.~Rizos, K.~Tamvakis and E.~Winstanley,
Phys. Rev. D \textbf{54} (1996), 5049-5058
doi:10.1103/PhysRevD.54.5049 [arXiv:hep-th/9511071 [hep-th]].



\bibitem{Kanti:1997br}
P.~Kanti, N.~Mavromatos, J.~Rizos, K.~Tamvakis and E.~Winstanley,
Phys. Rev. D \textbf{57} (1998), 6255-6264
doi:10.1103/PhysRevD.57.6255 [arXiv:hep-th/9703192 [hep-th]].







\bibitem{GBM:2017lvd}
  B.~P.~Abbott {\it et al.}
  ``Multi-messenger Observations of a Binary Neutron Star Merger,''
  Astrophys.\ J.\  {\bf 848} (2017) no.2,  L12
  doi:10.3847/2041-8213/aa91c9
  [arXiv:1710.05833 [astro-ph.HE]].



\bibitem{Akrami:2018odb}
  Y.~Akrami {\it et al.} [Planck Collaboration],
  arXiv:1807.06211 [astro-ph.CO].




\bibitem{DeFelice:2011zh}
  A.~De Felice and S.~Tsujikawa,
  JCAP {\bf 1104} (2011) 029
  doi:10.1088/1475-7516/2011/04/029
  [arXiv:1103.1172 [astro-ph.CO]].







\end{thebibliography}
\end{document}